\documentclass[prb,aps,twocolumn,superscriptaddress]{revtex4-2}
\usepackage{graphicx}
\usepackage{dcolumn}
\usepackage{bm}
\usepackage{soul}
\usepackage{lipsum}
\usepackage{ragged2e}
\usepackage{braket}
\usepackage{bbm}
\usepackage{lipsum}
\usepackage{color}
\usepackage{xcolor}
\usepackage{amsfonts}
\usepackage{amsmath}
\usepackage{amsmath,amssymb,latexsym}
\usepackage{wasysym,stmaryrd,marvosym}
\usepackage{mathrsfs}
\usepackage{soul}
\usepackage{dsfont}
\usepackage{float}

\usepackage[mathlines]{lineno}
\usepackage[colorinlistoftodos]{todonotes}
\usepackage[colorlinks=true, allcolors=blue]{hyperref}

\begin{document}

\newcommand{\ohio}{Department of Physics and Astronomy and Nanoscale and Quantum Phenomena Institute, Ohio University, Athens, Ohio 45701}

\title{Superconductivity and geometric superfluid weight of a tunable flat band system}

\author{M. A. Mojarro}
\email{mm232521@ohio.edu}
\author{Sergio E. Ulloa}
\affiliation{\ohio}

\date{\today}

\begin{abstract}
We study superconductivity and superfluid weight of the two-dimensional $\alpha$-$\mathcal{T}_3$ lattice with on-site asymmetries, 
hosting an isolated quasi-flat band with tunable bandwidth via a parameter $\alpha$.
Within a mean-field approximation of the attractive Hubbard model, we obtain the superconducting order parameters on the three inequivalent sublattices and show their strong dependence on $\alpha$, interaction strength, and electron filling.
At quasi-flat band filling, 
a superconducting gap opens and grows power-law fast with interaction strength, 
instead of the usual slow exponential growth, due to diverging density of states. 
We calculate the superfluid weight from linear response theory and study its band dispersion and geometric contributions. 
While the conventional part proportional to band derivatives is suppressed in the quasi-flat band regime, 
the contribution dominated by the quantum metric grows linearly for small interaction strength. 
We further demonstrate how tuning $\alpha$ enhances the quantum metric and thus the geometric superfluid weight especially near half-filling, 
while increasing on-site asymmetries increases the conventional contribution by broadening the quasi-flat band. 
We obtain the Berezinskii-Kosterlitz-Thouless transition temperature and demonstrate its strong dependence and enhancement with the parameter $\alpha$.
Our results establish a tunable flat band system, the $\alpha$-$\mathcal{T}_3$ lattice model, as a candidate for tunable quantum geometry and superfluid weight 
and as a prototype of related 
behavior in tunable quantum materials.

\end{abstract}


\maketitle

\section{Introduction}

Quantum geometry in solid state systems has emerged as a fundamental framework to describe physical phenomena governed by the evolution of Bloch wave functions in momentum space \cite{Liu2024,yu2025,gao2025,Kang2025}.
The 
quantum geometric tensor (QGT) 
\cite{Provost1980, Resta2011}, that specifies distances between Bloch states,
is composed of two fundamental quantities: an imaginary part, known as the Berry curvature \cite{Berry} and widely used to describe topological phases of matter \cite{Hasan2010, Qi2011}, and its real part, the quantum metric \cite{Provost1980}.

While the role of the Berry curvature in solids is well-understood, 
the quantum metric has remained unexplored until recently, and it is becoming a fundamental quantity for the understanding of physical responses 
involving multi-orbital mixing \cite{Essay, 2verma2025},
nonlinear transport phenomena \cite{Ahn2022, Anyuan2023,Wang2023} ,
and capacitance of insulators, among others \cite{Komissarov2024,verma2025,Yahir2025}.
In multiband systems, flat band superconductivity is enabled by the quantum metric \cite{Peotta2015,Liang2017,peotta2023}.
The quantum metric has been shown to enhance the superconducting transition temperature, 
making quantum geometry an interesting source of high-$T_c$ values.
Moreover, in two dimensions, the Berezinskii-Kosterlitz-Thouless (BKT) transition temperature depends strongly on the quantum geometry of the narrow bands in twisted bilayer graphene, whose geometric contribution to the superfluid weight has been probed experimentally \cite{Hu2019,Julku2020,Tanaka2025}.

The geometric superfluid weight has been explored
in flat band models such as Lieb \cite{Julku2016,Penttila2025} and dice lattices \cite{Wu2021}, as well as
other systems with tunable quantum metric \cite{Thumin2023,thumin2025}.
An interesting lattice hosting a flat band 
is the $\alpha$-$\mathcal{T}_3$ model \cite{Raoux2014,Illes2015}. This 
consists of a honeycomb lattice with an additional site at the center of each hexagon that connects only to one of the honeycomb sublattices. 
Its energy spectrum consists of graphene-like bands hosting Dirac cones at the corners of the Brillouin zone, with an additional flat band at zero energy.
The parameter $\alpha$ controls the wavefunction delocalization through relative hopping matrix elements,
while the energy spectrum remains $\alpha$-independent. 
The lattice interpolates between the honeycomb lattice ($\alpha=0$), and the dice lattice ($\alpha=1$), with different transport properties as a consequence of this interpolation \cite{Mojarro2020,Dey2019,Illes2017,Illes2016}.
Interestingly, as the wavefunctions do depend on the $\alpha$ parameter, an $\alpha$-dependent Berry phase leads to tunable electromagnetic properties in this model \cite{Raoux2014,Illes2015,Dey2018}. 
Although the physics resulting from the well-known dispersive bands in graphene has been studied in numerous works \cite{graphene2009}, 
the flat band in the dice lattice limit has only been recently observed in the electride YCl through  ARPES measurements \cite{geng2025}. 

In this work, we study superconductivity and superfluid weight of the $\alpha$-$\mathcal{T}_3$ lattice with on-site asymmetries.
The on-site asymmetries make the energy spectrum $\alpha$-dependent, and the otherwise flat band acquires a tunable finite width.
In a BCS mean-field theory, we calculate self-consistently the superconducting order parameters at the three sublattices, 
and the subsequent superfluid weight.
After the separation of geometric and conventional components, 
we find a highly tunable geometric contribution to the superfluid weight due to the strong $\alpha$-dependence of the quantum metric belonging to the quasi-flat band.
Finally, we study the dependence on $\alpha$ of the BKT transition temperature in this two-dimensional system.

The manuscript is organized as follows. In Sec.\,\ref{sec2} we present the model Hamiltonian for the $\alpha$-$\mathcal{T}_3$ lattice with on-site asymmetries. 
In Sec.\,\ref{sec3} we present the mean-field theory formalism for the superconducting order parameters, and study the superfluid weight in Sec.\,\ref{sec4}.
The finite temperature superfluid weight is studied in Sec.\,\ref{sec5} along with estimates of the BKT transition temperature as a function of the $\alpha$ parameter.
Finally, in Sec.\,\ref{sec6} we present our conclusions.




\section{$\alpha$-$\mathcal{T}_3$ lattice model with on-site asymmetries}\label{sec2}

\begin{figure}
    \centering
    \includegraphics[scale=0.32]{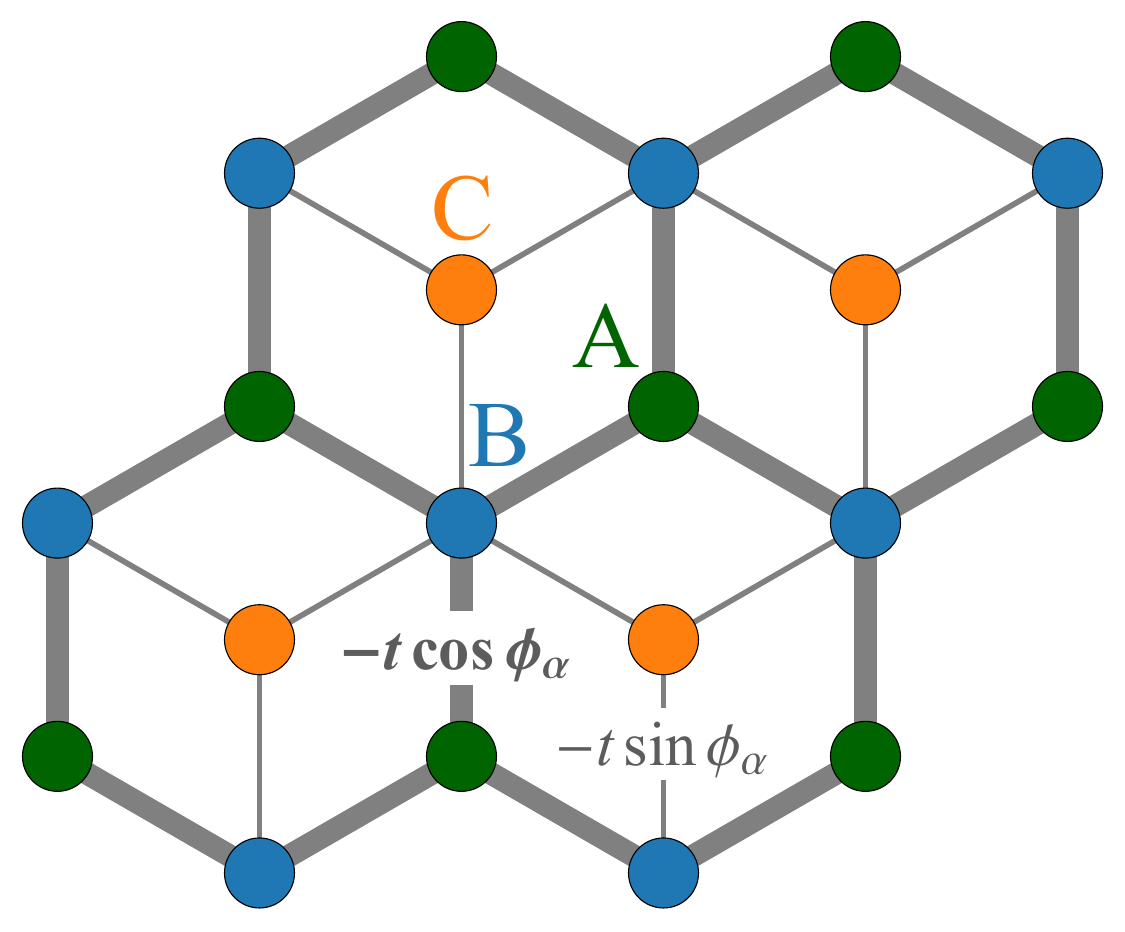}
    \caption{The $\alpha$-$\mathcal{T}_3$ lattice connects sublattices A and B with electron hopping energy  $-t\cos\phi_\alpha$, and sublattices B and C with hopping energy  $-t\sin\phi_\alpha$, where $\alpha=\tan\phi_\alpha$ and $\alpha\in(0,1]$. 
    For small $\alpha$, the hopping amplitude between A and B sites dominates. As $\alpha\rightarrow0$, the C sites become disconnected.}
    \label{lattice}
\end{figure}

The $\alpha$-$\mathcal{T}_3$ lattice 
has a triangular crystal structure with primitive vectors ${\bf a}_1=(\sqrt{3},\,0)$ and ${\bf a}_2=(\sqrt{3}/2,\,3/2)$, 
and three sites (A, B, and C) per unit cell.
The A and B sublattices form a honeycomb lattice with sites connected by nearest-neighbor vectors: $\bm{\delta}_1=(\sqrt{3}/2\,,1/2)$, $\bm{\delta}_2=(-\sqrt{3}/2\,,1/2)$, $\bm{\delta}_3=(0\,,-1)$. 
The C sites lie in the center of each hexagon and are connected only to B sites (see Fig.\,\ref{lattice}). 
Notice that the B and C sublattices form a staggered honeycomb lattice too.

In a single-orbital tight-binding model of hopping electrons, the hopping energy between different sublattices is tuned via a parameter $\alpha$, and described by the following spinless Hamiltonian \cite{Piechon2015}
\begin{eqnarray}
\label{H0}\mathcal{H}&=&\mathcal{H}_0+\mathcal{H}_\varepsilon\,,\\
\nonumber\label{H1}\mathcal{H}_0&=&-t\sum_{\bf{r}}\sum_{l=1}^3\left(b_{\textbf{r}}^{\dagger} a_{\textbf{r}+\bm{\delta}_l}\cos\phi_{\alpha} +  b_{\textbf{r}}^{\dagger} c_{\textbf{r}-\bm{\delta}_l}\sin\phi_\alpha \right) + \text{H.c.},\\\\
\label{H2}\mathcal{H}_\varepsilon&=&\sum_{\bf{r}}\left(\varepsilon_{\text{A}}a^\dagger_{\textbf{r}+\bm{\delta}_1}a_{\textbf{r}+\bm{\delta}_1}+\varepsilon_{\text{B}}b^\dagger_{\textbf{r}}b_{\textbf{r}}+\varepsilon_{\text{C}}c^\dagger_{\textbf{r}-\bm{\delta}_1}c_{\textbf{r}-\bm{\delta}_1}\right)\,,
\end{eqnarray}
where $a_{\textbf{r}+\bm{\delta}_l}$, $b_{\bf r}$, and $c_{\textbf{r}-\bm{\delta}_l}$ ($a^\dagger_{\textbf{r}+\bm{\delta}_l}$, $b^\dagger_{\bf r}$, and $c^\dagger_{\textbf{r}-\bm{\delta}_l}$) are annihilation (creation) operators associated to sublatices A, B, and C, respectively, and $\bf r$ runs over B sites.
The hopping energy $-t\cos\phi_\alpha$ between B and A sites, and $-t\sin\phi_\alpha$ between B and C sites are tuned with a parameter $\alpha\in(0,1]$  through $\alpha=\tan\phi_\alpha$.
Notice that A and C sites are not connected, and that the model is dual to the change $\alpha\rightarrow1/\alpha$. We have also included on-site energies $\varepsilon_i$ ($i=$A, B, C) to account for possible different atomic species or chemical environment.

In the limit $\alpha\rightarrow0$, the C sites become disconnected from B sites and the Hamiltonian describes a graphene-like system of A and B sites. 
On the other hand, when $\alpha=1$, the dice lattice is recovered with all nearest-neighbor hopping energies being the same, so that the lattice enjoys $C_{6v}$ symmetry and the A and C sublattices become equivalent.
In contrast, when $\alpha\neq1$, $C_{6v}$ is broken down to $C_{3v}$ as A and C sites become inequivalent.
The second term in the Hamiltonian, $H_\varepsilon$, accounts for sublattice-dependent on-site energies
that determine the energy separation between bands.
We consider symmetric on-site energies with respect to
the C sites, 
so that $\varepsilon_{\text{C}}=0$, and $\varepsilon_{\text{A}}=-\varepsilon_{\text{B}}=\varepsilon$;
in this case, the energy spectrum contains an isolated nearly flat band close to zero energy, as shown below. 
Different on-site asymmetry configurations lead to quantitatively similar results.

\begin{figure}
    \centering
    \includegraphics[scale=0.45]{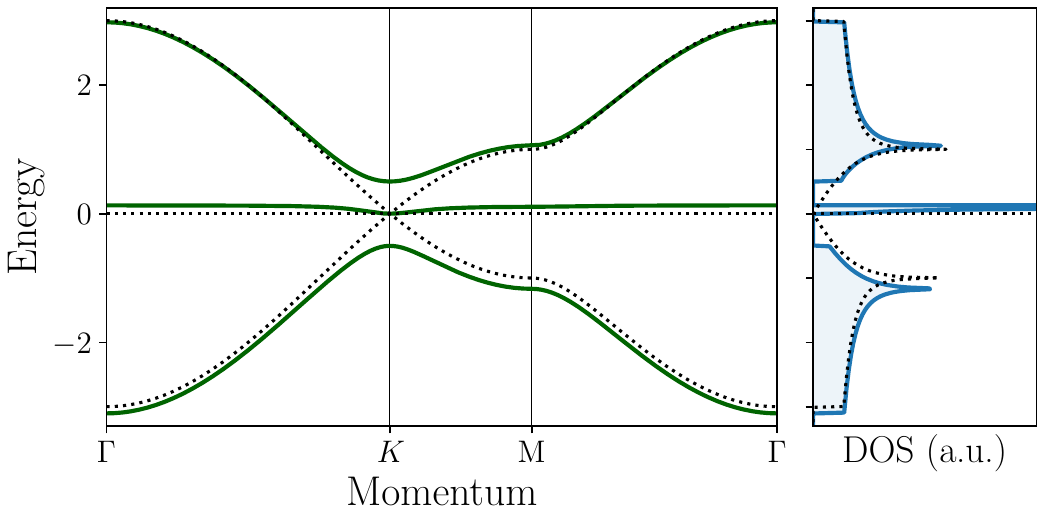}
    \caption{Energy bands and density of states of the $\alpha$-$\mathcal{T}_3$ lattice with $\alpha=0.6$ and $\varepsilon=0.5$ along a high-symmetry path in the Brillouin zone. Dotted lines correspond to $\varepsilon=0$, where the energy spectrum becomes $\alpha$-independent.}
    \label{bandas}
\end{figure}

\begin{figure}
    \centering
    \includegraphics[scale=0.65]{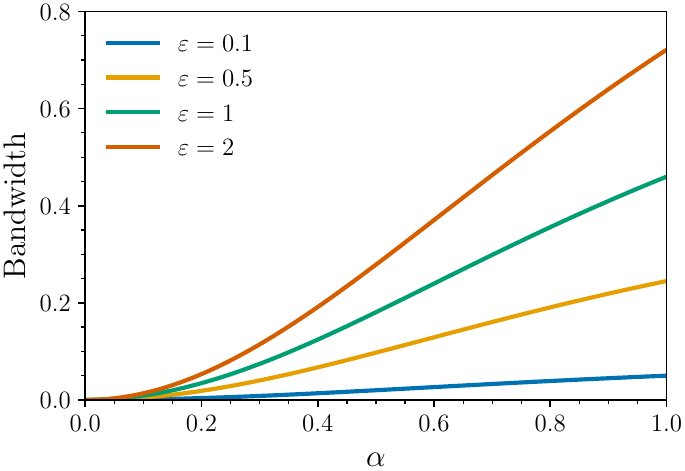}
    \caption{Bandwidth of the quasi-flat band $\varepsilon_1({\bf k})$ ($i=1$ in Eq.\,\eqref{bands}) as a function of $\alpha$ for several values of A-B site asymmetry $\varepsilon$.}
    \label{BW}
\end{figure}

We can write the Hamiltonian \eqref{H0} in momentum space as $\mathcal{H}=\sum_{\bf k}\psi_{\bf k}^{\dagger}H({\bf k})\psi_{\bf k}$, where $\psi_{\bf k}=(a_{{\bf k}},\,b_{{\bf k}},\,c_{{\bf k}})^{\text{T}}$ contains annihilation operators, and the Bloch Hamiltonian $H({\bf k})$ is \cite{Raoux2014,Illes2015,Dey2018,Mojarro2020},
\begin{equation}\label{Ham}
    H({\bf k})=\begin{pmatrix}
    \varepsilon & g({\bf k})\cos\phi_\alpha & 0\\
    g^\ast({\bf k})\cos\phi_\alpha & -\varepsilon &g({\bf k})\sin\phi_\alpha\\
    0 & g^\ast({\bf k})\sin\phi_\alpha & 0
    \end{pmatrix}\,,
\end{equation}
where we have defined $g({\bf k})=-t\sum_{i=1}^3e^{-i{\bf k}\cdot\bm{\delta}_i}$, and ${\bf k}=(k_x,\,k_y)$ is the electron wave-vector. 
In the following, we take $t=1$ as energy unit. The Hamiltonian eigenvalues are \cite{Dey2019}
\begin{equation}\label{bands}
    \varepsilon_i({\bf k})=2\sqrt{\frac{p({\bf k})}{3}}\cos\left[\frac{1}{3}\cos^{-1}\left(-\frac{3q({\bf k})}{2p({\bf k})}\sqrt{\frac{3}{p({\bf k})}}\right)-\frac{2\pi}{3}i\right]\,,
\end{equation}
where $i=0,1,2$.
We have defined $p({\bf k})=|g({\bf k})|^2+\varepsilon^2$, and $q({\bf k})=\varepsilon|g({\bf k})|^2\sin^2\phi_\alpha$.
When $\varepsilon=0$, the energy spectrum becomes $\alpha$-independent
and one recovers $\varepsilon_0({\bf k})=|g({\bf k})|$ and $\varepsilon_2({\bf k})=-|g({\bf k})|$, describing conduction and valence graphene-like bands with Dirac cones at the corners of the Brillouin zone, respectively. 
In addition, a flat band at zero energy $\varepsilon_1=0$ (see dotted lines in Fig.\,\ref{bandas}) results from destructive quantum interference of hopping electrons \cite{Leykam01012018}.
In this limit, the eigenstates corresponding to $\varepsilon_0({\bf k})$ and $\varepsilon_2({\bf k})$ are $\ket{\psi_\pm}=(\cos\phi_{\alpha}e^{i\theta({\bf k})},\,\pm 1,\,\sin\phi_\alpha e^{-i\theta({\bf k})})^{\text{T}}/\sqrt{2}$, respectively, where $\theta({\bf k})$ is the phase defined by $g({\bf k})=|g({\bf k})|e^{i\theta({\bf k})}$  \cite{Illes2015}.
The flat band has eigenstate $\ket{\psi_\text{FB}}=(\sin\phi_{\alpha}e^{i\theta({\bf k})},\,0 ,\,-\cos\phi_\alpha e^{-i\theta({\bf k})})^{\text{T}}$.
We observe that when $\alpha\rightarrow0$ ($\phi_{\alpha}\rightarrow0$), the states of the conduction and valence bands spread over A and B sites exclusively, while the state of the flat band becomes localized at C sites.
In this limit, there is no hopping between B and C sites, and the flat band is of atomic type. 
A finite $\varepsilon$ opens a gap at the Dirac point between the three bands and the energy spectrum becomes $\alpha$-dependent (see Fig.\,\ref{bandas}). 
Increasing $\alpha$ or $\varepsilon$ leads to an increasing width of the quasi-flat band as shown in Fig.\,\ref{BW}. 
The tunability of this quasi-flat band and its associated eigenstate as a function of $\alpha$  has strong effects on the superfluid weight and quantum geometry, as we discuss below.

\section{superconducting order parameters}\label{sec3}

To explore the superfluid weight in the $\alpha$-$\mathcal{T}_3$ lattice, we calculate the superconducting order parameters in a mean-field formalism. 
We consider the following attractive Hubbard Hamiltonian 
\begin{eqnarray}
    \label{Htotal}\mathcal{H}&=&\mathcal{H}_0+\mathcal{H}_\varepsilon+\mathcal{H}_U+\mathcal{H}_\mu\,,
    \\\mathcal{H}_U&=&-U\sum_{i} f^\dagger_{i\uparrow}f^\dagger_{i\downarrow}f_{i\downarrow}f_{i\uparrow}\,,\\
    \mathcal{H}_\mu&=&-\mu\sum_i \left(f^\dagger_{i\uparrow}f_{i\uparrow}+f^\dagger_{i\downarrow}f_{i\downarrow} \right)\,,
\end{eqnarray}
where the fermionic operator $f_{i\sigma}$ annihilates an electron with spin $\sigma=\uparrow,\downarrow$ at site $i$. Here $U>0$ is the interaction energy and $\mu$ the chemical potential introduced to control band filling. $\mathcal{H}_0$ and
$\mathcal{H}_\varepsilon$ are defined in Eqs.\,\eqref{H1} and \eqref{H2}, respectively. 
In a mean-field theory, we approximate $f_{i\downarrow}f_{i\uparrow}\approx-\Delta_i/U+\delta$, where $\delta$ is a small fluctuation around the mean value (order parameter) $\Delta_i=-U\braket{f_{i\downarrow}f_{i\uparrow}}$. 
Keeping only first-order terms in $\delta$, we decouple the interaction term as $f^\dagger_{i\uparrow}f^\dagger_{i\downarrow}f_{i\downarrow}f_{i\uparrow} \approx -(\Delta_i/U) f_{i\uparrow}^{\dagger}f_{i\downarrow}^{\dagger}-(\Delta_i/U) f_{i\downarrow}f_{i\uparrow}-\Delta_i^2/U^2$, where we have chosen $\Delta_i$ to be real.
We consider sublattice dependent pairing, so that there is a total of three superconducting order parameters, namely $\Delta_\text{A}$, $\Delta_\text{B}$, and $\Delta_\text{C}$.

Introducing particle-hole symmetry, we can write the Hamiltonian \eqref{Htotal} in momentum space as the following Bogoliubov-de-Gennes (BdG) form
\begin{equation}\label{HBDG}
    H_{\text{BdG}}({\bf k})=\begin{pmatrix}
        H_{\uparrow}({\bf k})-\mu & \Gamma\\
        \Gamma^\dagger & -H_{\downarrow}^\ast(-{\bf k})+\mu
    \end{pmatrix}\,,
\end{equation}
where $\Gamma=\text{diag}(\Delta_\text{A},\,\Delta_\text{B},\,\Delta_\text{C})$. In our case, as time-reversal symmetry is preserved, 
we have that $H_{\uparrow}({\bf k})=H^\ast_{\downarrow}(-{\bf k})=H(\bf k)$, given in Eq.\,\eqref{Ham}. 
Imposed by particle-hole symmetry, $H_{\text{BdG}}({\bf k})$ has eigenvalues $\pm E_i({\bf k})$ ($i=1,2,3$), which are calculated numerically.

The order parameters in the ground-state are found by minimizing the mean-field free energy $\mathcal{F}=\Omega+\mu N_e$, where $N_e$ is the number of electrons and $\Omega=-2/\beta\sum_i\sum_{\bf k}\ln[2\cosh(\beta E_i({\bf k})/2)]+(\Delta_{\text{A}}^2+\Delta_{\text{B}}^2+\Delta_{\text{C}}^2)/(3U)-\mu N$ is the grand canonical potential, 
where $\beta=1/k_BT$ is the inverse temperature and $N$ the number of sites.
We solve the system of equations $\partial\mathcal{F}/\partial\Delta_i=0$ for $i=\text{A},\text{B},\text{C}$ self-consistently as function of $U$, together with $N_e=-\partial\Omega/\partial\mu$
in order to control the chemical potential for a fixed electron density per site $n_e=N/N_e$.
In the following we take the zero temperature limit $\beta\rightarrow\infty$ to characterize the ground state as a function of $\alpha$, $\varepsilon$ and $U$.

In the symmetric dice-lattice case, $\alpha=1$ and $\varepsilon=0$, we find that $\Delta_{\text{A}}=\Delta_{\text{C}}\neq\Delta_{\text{B}}$ for any electron interaction $U$, as the A and C sublattices become indistinguishable in this limit due to the $C_{6v}$ symmetry of the lattice. 
In contrast, in the graphene limit $\alpha\rightarrow0$, we obtain $\Delta_{\text{A}}=\Delta_{\text{B}}$ following the known graphene behavior as a function of $U$ \cite{Zhao2006}. 
The $\Delta_{\text{C}}$ parameter saturates in this limit for large enough $U$ due to local pair formation, but does not participate in the formation of a superconducting state due to the absence of hopping to the C sublattice.

\begin{figure}
    \centering
    \includegraphics[scale=0.65]{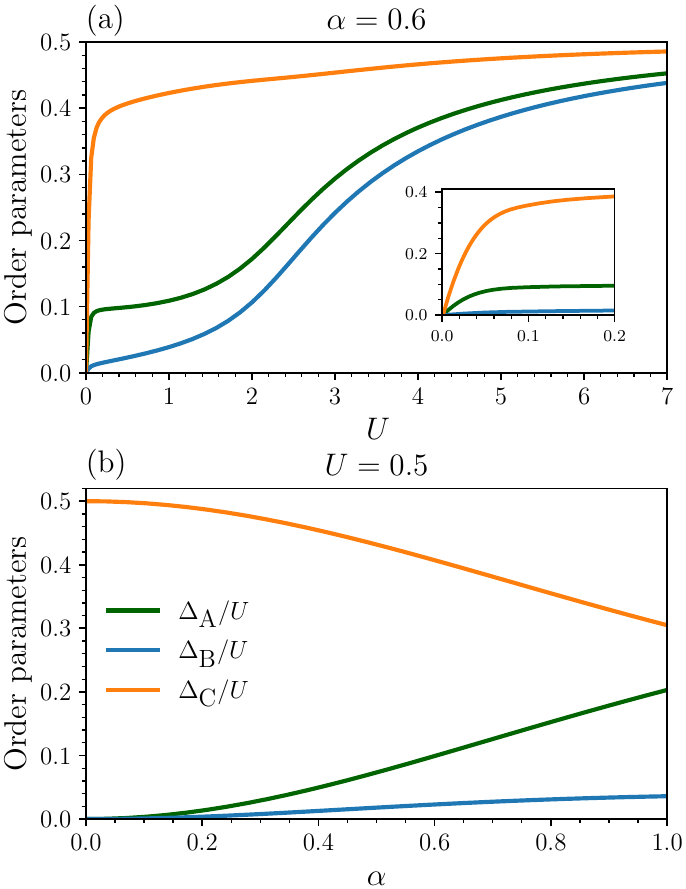}
    \caption{(a) Zero temperature superconducting order parameters $\Delta_i/U$ as a function of $U$ for $\alpha=0.6$. The inset shows the behavior at small $U$. In (b), we set $U=0.5$ and plot as a function of $\alpha$. In both cases $\varepsilon=0.5$ and $n_e=1$ (half-filling).}
    \label{DeltaABC}
\end{figure}

For $0<\alpha<1$ and $\varepsilon\neq0$, the three order parameters become distinct as the three sublattices are inequivalent. 
In the range of quasi-flat band filling, $n_e\in(2/3,\,4/3)$, the $U$ onset for nonzero superconducting pairing becomes vanishingly small, and the magnitude of the order parameters increase rapidly with $U$; they do not follow the usual exponentially small $U$ dependence due to the diverging density of states \cite{Kopnin2011}.
Away from quasi-flat band filling, the order parameters follow an usual exponentially small U dependence.
Fig.\,\ref{DeltaABC}(a) shows the ground state order parameters as a function of $U$ for $\alpha=0.6$ and $\varepsilon=0.5$ at half-filling ($n_e=1$).
We observe that nonzero pairing is achieved for $U\gtrsim0$, 
and that $\Delta_{\text{C}}$ is larger overall as Cooper pairs tend to localize at C sites due to their isolated nature and 
the associated quasi-flat band wavefunction localization on this sublattice. Similarly, note that $\Delta_\text{A}>\Delta_\text{B}$, as the quasi-flat band wavefunction has a larger weight on the A sublattice than on the B sublattice.

To investigate the effect of a varying $\alpha$, Fig.\,\ref{DeltaABC}(b) shows the superconducting order parameters as a function of $\alpha$ for $U=0.5$ at half-filling. 
We observe how $\Delta_{\text{A}}$ and $\Delta_{\text{C}}$ get closer in magnitude as $\alpha$ increases, as both sublattices become more equivalent with increasing $\alpha$.
The difference between $\Delta_{\text{A}}$ and $\Delta_{\text{C}}$ when $\alpha=1$ is due to the finite on-site asymmetry $\varepsilon$.
Again, the parameter $\Delta_{\text{B}}$ is smaller as the quasi-flat band wavefunction has a vanishingly small weight in such sublattice.
The overall larger magnitude of $\Delta_{\text{C}}$ is again understood from its isolated nature.

As a superconducting state is defined when a nonzero superfluid stiffness develops in the system, in the next section we calculate the superfluid weight for several system parameters and explore the role of quantum geometry.

\section{Superfluid weight}\label{sec4}

In linear response theory, the current density induced in a superconductor due to an electromagnetic gauge field ${\bf A}({\bf r},\,t)$ is given in Fourier space by $J_i({\bf q},\,\omega)=-\sum_jK_{ij}({\bf q},\,\omega)A_j({\bf q},\,\omega)$ \cite{mahan2013many}, where 
$K_{ij}({\bf q},\,\omega)=\braket{T_{ij}}-i\int_0^{\infty}dt\,e^{i\tilde{\omega} t}\braket{[j_i({\bf q},\,t),\,j_j(-{\bf q},\,0)]}$ is the current-current response function, where the brackets $\braket{\,}$ denote a mean value with respect to the BdG Hamiltonian.
Here $\omega$ and $\bf q$ are the frequency and momentum of the external field, respectively. 
We have also defined $\tilde{\omega}=\omega+i0^+$ to achieve convergence.
The first term accounts for the diamagnetic contribution with $T_{ij}=\sum_{\bf k}\psi_{{\bf k}\sigma}^{\dagger}\partial_i\partial_jH_{\sigma}({\bf k})\psi_{{\bf k}\sigma}$,
while the second term includes the paramagnetic current operator determined by
$j_i({\bf q})=\sum_{{\bf k}\sigma}\psi^\dagger_{{\bf k}\sigma}\partial_iH_{\sigma}({\bf k}+{\bf q}/2)\psi_{{\bf k}+{\bf q}\sigma}$. 
Here we have used the compact notation $\partial_i\equiv\partial/\partial k_i$.
The superfluid weight $D^s_{ij}$ is defined as the zero-momentum limit of the static current-current response function, that is $D^s_{ij}=\lim_{{\bf q}\rightarrow0}K_{ij}({\bf q},\,0)$ \cite{Scalapino1993}. 
In Ref.\,\cite{Liang2017}, it was shown that for local pairing interactions, the superfluid weight acquires the following form
\begin{eqnarray}
    \nonumber    D_{ij}^s&=&\sum_{\lambda\lambda'}\int_{\text{BZ}}\frac{d^2k}{(2\pi)^2}\frac{n(E_{\lambda'}({\bf k}))-n(E_{\lambda}({\bf k}))}{E_{\lambda}({\bf k})-E_{\lambda'}({\bf k})}\\
    \nonumber&&\times\big(\braket{\Psi_\lambda|\partial_i H_{{\text{BdG}}}({\bf k})|\Psi_{\lambda'}}\braket{\Psi_{\lambda'}|\partial_j H_{{\text{BdG}}}({\bf k})|\Psi_{\lambda}}\\
    \nonumber&&-\braket{\Psi_\lambda|\partial_i H_{{\text{BdG}}}({\bf k})\gamma_z|\Psi_{\lambda'}}\braket{\Psi_{\lambda'}|\partial_j H_{{\text{BdG}}}({\bf k})\gamma_z|\Psi_{\lambda}}\big)\,,\\
\end{eqnarray}
where $n(E)=1/(e^{E\beta}+1)$ is the Fermi-Dirac distribution function, and $\gamma_z$ is a Pauli matrix acting on particle-hole space.
$E_{\lambda}({\bf k})$ and $\ket{\Psi_\lambda}$ are the eigenvalues and eigenvectors of the BdG Hamiltonian in Eq.\,\eqref{HBDG}, respectively.
In order to reveal the quantum geometric properties of the superfluid weight, 
we write the BdG states as $\ket{\Psi_\lambda}=\sum_m a_{m\lambda}\ket{m}_p+ b_{m\lambda}\ket{m}_h$,
where $\ket{m}_p=\ket{m}\otimes\ket{p}$, and $\ket{m}_h=\ket{m}\otimes\ket{h}$, with $\ket{m}$ an eigenvector of $H({\bf k})$ in \eqref{Ham}, 
and $\ket{p}$ ($\ket{h}$) the eigenvector of $\gamma_z$ with eigenvalue $+1$ ($-1$). 
Using this expansion we recover
\begin{eqnarray}
    \nonumber    D_{ij}^s&=&2\sum_{\lambda\lambda'}\sum_{mnqp}\int_{\text{BZ}}\frac{d^2k}{(2\pi)^2}G_{\lambda\lambda'}({\bf k})v_{i}^{mn}v_{j}^{qp}\\
    \nonumber&&\qquad\qquad\times\big(a_{m\lambda}^\ast a_{n\lambda'}b_{q\lambda'}^\ast b_{p\lambda}+a_{q\lambda'}^\ast a_{p\lambda}b_{m\lambda}^\ast b_{n\lambda'}\big)\,,\\
\end{eqnarray}
where $v_i^{mn}=\braket{m|\partial_iH({\bf k})|n}=\partial_i\varepsilon_n({\bf k})\delta_{mn}+(\varepsilon_n({\bf k})-\varepsilon_m({\bf k}))\braket{m|\partial_i n}$ are the velocity matrix elements, and we have defined $G_{\lambda\lambda'}({\bf k})=[n(E_{\lambda}({\bf k}))-n(E_{\lambda'}({\bf k}))]/[E_{\lambda}({\bf k})-E_{\lambda'}({\bf k})]$.
In our case we have $D_{ij}^s=D^s\delta_{ij}$ with
\begin{eqnarray}
    \nonumber    D^s&=&4\sum_{\lambda\lambda'}\sum_{mnqp}\int_{\text{BZ}}\frac{d^2k}{(2\pi)^2}G_{\lambda\lambda'}({\bf k})v_{i}^{mn}v_{i}^{qp}a_{m\lambda}^\ast a_{n\lambda'}b_{q\lambda'}^\ast b_{p\lambda}\,.
\end{eqnarray}
\begin{figure}
    \centering
    \includegraphics[scale=0.7]{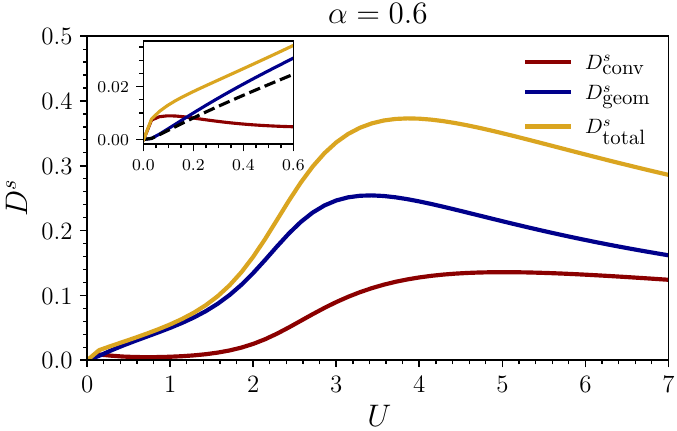}
    \caption{Superfluid weight as a function of $U$ for $\alpha=0.6$ at half-filling ($n_e=1$).  The inset shows the small $U$ regime, where the geometric contribution grows linearly with $U$. Dashed line shows the contribution from the quantum metric in Eq.\,\eqref{Dgeomg}. Here we take $\varepsilon=0.5$.}
    \label{SWalp06}
\end{figure}
We can decompose the superfluid weight in terms of its conventional and geometric contributions \cite{Julku2016,Liang2017}
\begin{equation}\label{totalD}
    D^s=D_{\text{conv}}^{s}+D_{\text{geom}}^{s}\,,
\end{equation}
where the conventional term includes exclusively intraband components ($m=n$, $q=p$) and is proportional to band dispersion derivatives as
\begin{eqnarray}
    \nonumber    D^{s}_{\text{conv}}&=&4\sum_{\lambda\lambda'}\sum_{mq}\int_{\text{BZ}}\frac{d^2k}{(2\pi)^2}G_{\lambda\lambda'}({\bf k}) \partial_i\varepsilon_m({\bf k})\partial_i\varepsilon_q({\bf k})\\&&\qquad \qquad \qquad \qquad \times a_{m\lambda}^\ast a_{m\lambda'}b_{q\lambda'}^\ast b_{q\lambda}\,,
\end{eqnarray}
while the geometric contribution $D^s_{\text{geom}}=D^s_{\text{geom},1}+D^s_{\text{geom},2}+D^s_{\text{geom},3}$ includes derivatives of the eigenstates and is determined by
\begin{eqnarray}
    \nonumber    D^{s}_{\text{geom,1}}&=&4\sum_{\lambda\lambda'}\sum_{\substack{m\neq n,q\neq p,\\(n\neq q\,\text{or}\, m\neq p)}}\int_{\text{BZ}}\frac{d^2k}{(2\pi)^2}G_{\lambda\lambda'}({\bf k})\\
    \nonumber&&\times(\varepsilon_n({\bf k})-\varepsilon_m({\bf k}))(\varepsilon_q({\bf k})-\varepsilon_p({\bf k}))\braket{m|\partial_i n}\braket{\partial_i q| p}\\
    \nonumber&&\times\, a_{m\lambda}^\ast a_{n\lambda'}b_{q\lambda'}^\ast b_{p\lambda}\,,\\
\end{eqnarray}
\begin{eqnarray}
    \nonumber    D^{s}_{\text{geom,2}}&=&4\sum_{\lambda\lambda'}\sum_{m\neq n,q}\int_{\text{BZ}}\frac{d^2k}{(2\pi)^2}G_{\lambda\lambda'}({\bf k})\\    \nonumber&&\times(\varepsilon_n({\bf k})-\varepsilon_m({\bf k}))\braket{m|\partial_i n}\partial_i\varepsilon_q({\bf k})\\
    &&\times\big(a_{m\lambda}^\ast a_{n\lambda'}b_{q\lambda'}^\ast b_{q\lambda}+a_{q\lambda}^\ast a_{q\lambda'}b_{m\lambda'}^\ast b_{n\lambda}\big)\,,
\end{eqnarray}
\begin{eqnarray}\label{Dgeomg}
    \nonumber    D^{s}_{\text{geom},3}&=&4\sum_{\lambda\lambda'}\sum_{m\neq n}\int_{\text{BZ}}\frac{d^2k}{(2\pi)^2}G_{\lambda\lambda'}({\bf k})\\
    \nonumber&&\times a_{m\lambda}^\ast a_{n\lambda'}b_{n\lambda'}^\ast b_{m\lambda}(\varepsilon_n({\bf k})-\varepsilon_m({\bf k}))^2g_{ii}^{nm}\,,\\
\end{eqnarray}
where we have used $\braket{q| \partial_j p}=-\braket{\partial_j q| p}$, and $g_{ij}^{nm}=\frac{1}{2}(\braket{\partial_in|m}\braket{m|\partial_jn}+\braket{\partial_jn|m}\braket{m|\partial_in})$ are the matrix elements of the quantum metric.
This quantity enters the quantum geometric tensor, which is defined for a band $n$ as $Q_{ij}^{n}=g_{ij}^{n}-\frac{i}{2}\Omega_{ij}^{n}$, where the real part is the band-resolved quantum metric $g_{ij}^{n}=\sum_{m\neq n}g_{ij}^{nm}$, and the imaginary part is the well-known Berry curvature $\Omega_{ij}^{n}=-2\,\text{Im}\braket{\partial_in|\partial_jn}$.
While the surface integral of the Berry curvature describes the phase accumulated by a state in an adiabatic closed loop in momentum space, the quantum metric captures the distance between states under infinitesimal changes in momentum \cite{Resta2011}. 
In the isolated band limit and uniform superconducting pairing, the geometric superfluid weight is determined by the term proportional to the quantum metric, $D^{s}_{\text{geom},3}$ \cite{Liang2017}.

\begin{figure}
    \centering
    \includegraphics[scale=0.63]{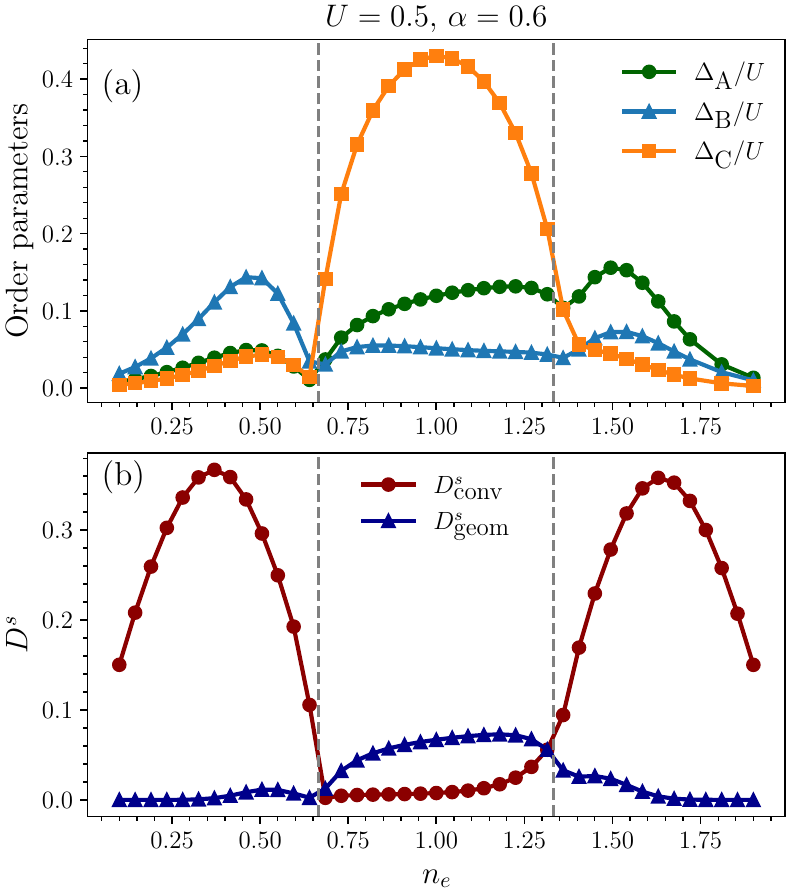}
    \caption{(a) Order parameters and (b) conventional and geometric superfluid weights as function of electron filling for $U=0.5$ and $\alpha=0.6$. The region between the dashed lines corresponds to quasi-flat band filling, where $D_{\text{geom}}^s>D_{\text{conv}}^s$. Here we take $\varepsilon=0.5$.}
    \label{SWne}
\end{figure}

In Fig.\,\ref{SWalp06} we show the superfluid weight as a function of $U$ for $\alpha=0.6$ at half-filling.
The inset shows how the total superfluid weight is governed by the geometric contribution for $U\gtrsim0.2$, especially $D^s_{\text{geom},3}$, involving the quantum metric of the quasi-flat band, as Cooper pairing occurs exclusively within this band \cite{Julku2020}.
Notice that for values of $U$ comparable to the bandwidth of the quasi-flat band or smaller, the conventional contribution dominates the geometric one, due to small but finite band derivatives.  
As $U$ increases, this contribution drops, and the geometric contribution grows linearly with $U$.
For larger $U\gtrsim2\varepsilon$, 
the conventional term grows due to the finite band derivatives of the graphene-like bands.
Outside the regime of quasi-flat band filling, the total superfluid weight is dominated by the conventional contribution (see Fig.\,\ref{SWne}),
as the derivatives of the dispersive bands become more relevant.

\begin{figure}
    \centering
    \includegraphics[scale=0.52]{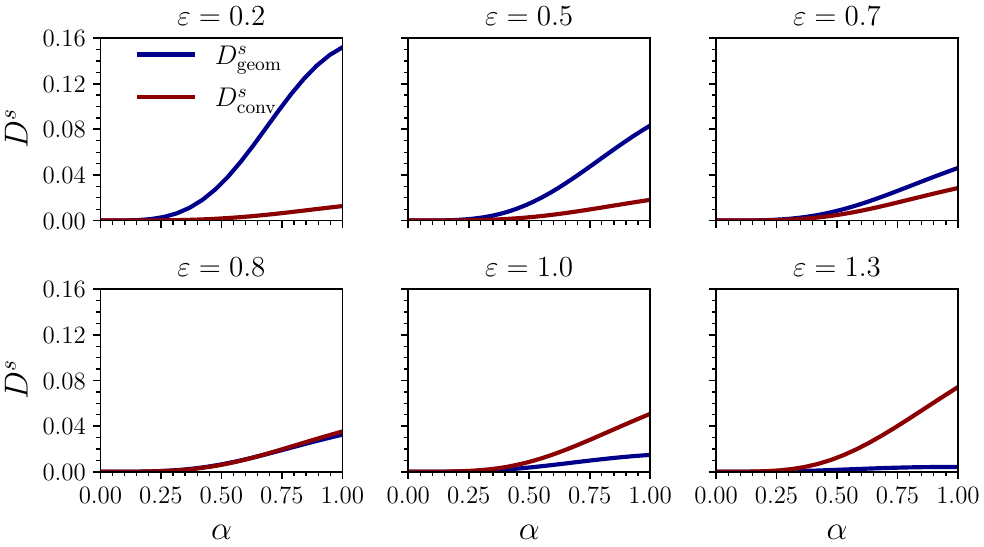}
    \caption{Conventional and geometric superfluid weight at half-filling for several values of $\varepsilon$ as a function of $\alpha$. We take $U=0.5$.}
    \label{Swalp}
\end{figure}

\begin{figure}
    \centering
    \includegraphics[scale=0.7]{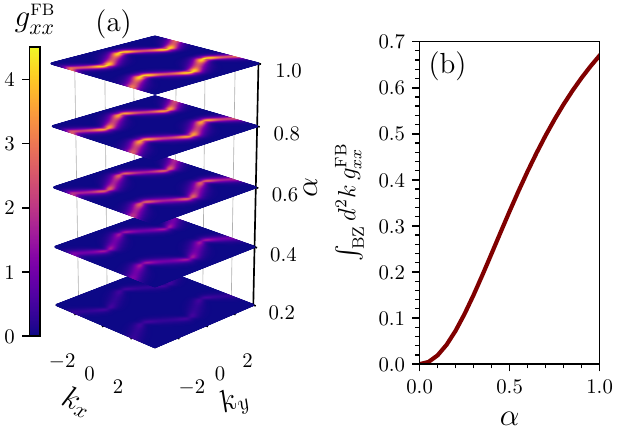}
    \caption{(a) Surface plot of the quantum metric of the flat band $g_{xx}^{\text{FB}}$ in momentum space for several values of $\alpha$. As $\alpha$ increases, the overall magnitude of $g^{\text{FB}}_{xx}$ increases as shown by its integral in (b).}
    \label{gxx}
\end{figure}

To investigate the dependence of the superfluid weight on the parameter $\alpha$ and on-site energy $\varepsilon$ at quasi-flat band filling, Fig.\,\ref{Swalp} shows the conventional and geometric superfluid weights at $U=0.5$ as a function of $\alpha$ for several values of $\varepsilon$.
We observe that as $\varepsilon$ increases, the conventional term involving band derivatives becomes more relevant as the bandwidth of the quasi-flat band grows with $\varepsilon$ (see Fig.\,\ref{BW}).
It is also observed that the geometric contribution grows with $\alpha$.
For this small value of $U$, Cooper pairing occurs predominantly within the quasi-flat band,
so that the contribution involving the quantum metric of the quasi-flat band dominates in $D^s_{\text{geom}}$ as mentioned previously.
Therefore, the increasing contribution as a function of $\alpha$ indicates an enhancement in the quantum metric.

This $\alpha$ dependence can be seen explicitly in Fig.\,\ref{gxx}(a), where the quantum metric of the quasi-flat band $g_{xx}^{\text{FB}}$ is shown for several values of $\alpha$ in momentum space.
We observe that the overall magnitude of $g_{xx}^{\text{FB}}$ increases with $\alpha$, also revealed in the integral of the quantum metric as a function of $\alpha$ in Fig.\,\ref{gxx}(b).
Notice that $g_{yy}^{\text{FB}}$ (not shown) has similar behavior, rotated by $\pi/2$ in momentum space. 

This behavior is understood as due to the changing spatial spread of the states in the quasi-flat band. As $\alpha$ increases, the state becomes more delocalized from the C sites, spreading over to the A sites. 
One can show that for small $\varepsilon$, $g_{ii}^{\text{FB}} \propto \sin^2(2\phi_\alpha)$. 
For $\phi_\alpha=0$, there is no spread of the quasi-flat band wave function, while for $\phi_\alpha=\pi/4$ the wave function is most delocalized and $g_{ii}^{\text{FB}}$ reaches maximum amplitude.


\begin{figure}
    \centering
    \includegraphics[scale=0.75]{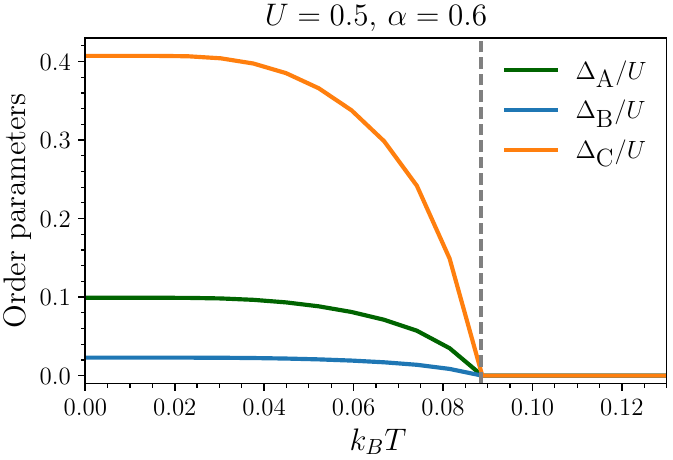}
    \caption{Superconducting order parameters as a function of temperature at half-filling, for $U=0.5$, $\alpha=0.6$ and $\varepsilon=0.5$. The vertical dotted line denotes the $T_{\text{BCS}}$ transition temperature, where the order parameters vanish.}
    \label{DeltT}
\end{figure}

\section{Berezinskii-Kosterlitz-Thouless transition temperature}\label{sec5}

We now study the temperature dependence of the order parameters and superfluid weight in the $\alpha$-$\mathcal{T}_3$ lattice. This requires calculating the superconducting order parameters self-consistently by minimizing the free energy, now at finite temperature at fixed $U$; we consider half-filling in what follows.

Cooper pair formation and gap opening in the BdG spectrum
are intrinsically connected to the existence of the BCS critical temperature $T_{\text{BCS}}$ in superconductors.
Fig.\,\ref{DeltT} shows the order parameters as a function of temperature at half-filling. 
The three order parameters in the $\alpha$-$\mathcal{T}_3$ lattice vanish for temperatures above $T_{\text{BCS}}$, where the Cooper pairs unbind.

\begin{figure}
    \centering
    \includegraphics[scale=0.75]{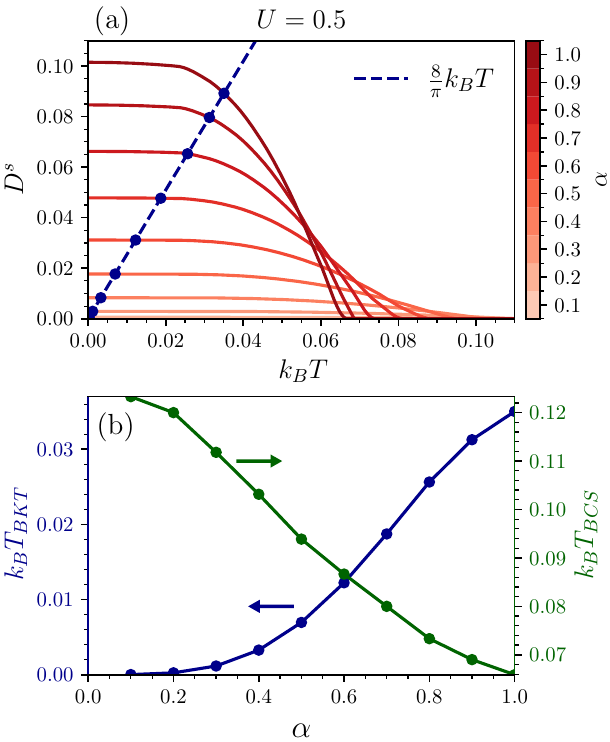}
    \caption{(a) Superfluid weight at half-filling as a function of temperature for several values of $\alpha$ for $U=0.5$ and $\varepsilon=0.5$. 
    The dotted line indicates $(8/\pi)k_BT$, where its intersection with $D^s(T)$ determines the BKT transition temperature.
    This temperature is shown in (b) as a function of $\alpha$, as well as $T_{\text{BCS}}$ where the order parameters vanish.}
    \label{Tbkt}
\end{figure}

On the other hand, it is known that in two-dimensional systems the superfluid phase develops below the Berezinskii-Kosterlitz-Thouless (BKT) temperature $T_{\text{BKT}}$ \cite{Kosterlitz1973}, 
which is lower than $T_{\text{BCS}}$ and related to the superfluid weight through the following equation \cite{Nelson1977}
\begin{equation}
    k_{B}T_{\text{BKT}}=\frac{\pi}{8}D^s(T=T_{\text{BKT}})\,.
\end{equation}
At $T>T_{\text{BKT}}$, vortices with positive and negative circulation unbind and coherence is lost, causing resistive behavior.
In the weak-coupling limit, when the flat band is partially filled, 
$T_{\text{BKT}}$ is enhanced and is determined by the quantum metric \cite{Liang2017}, since in this regime the superfluid weight is generally larger than in dispersive bands.
We are interested in studying how $T_{\text{BKT}}$ depends on the $\alpha$ parameter. 
The superfluid weight $D^s(T)$ for several values of $\alpha$ is shown in Fig.\,\ref{Tbkt}(a).
We observe a typical behavior of $D^s(T)$ decreasing as $T$ increases, until it vanishes at $T_{\text{BCS}}$.
The blue dotted line shows $(8/\pi)k_BT$, so that the intersection between this line and $D^s(T)$ determines the BKT transition temperature for each $\alpha$.  
Fig.\,\ref{Tbkt}(b) shows $T_{\text{BKT}}$ as a function of $\alpha$.
Notably, $T_{\text{BKT}}$ increases rapidly with $\alpha$, which is reminiscent of the enhancement of the quantum metric of the quasi-flat band as $\alpha$ grows.
In the same figure, we show the BCS transition temperature, displaying a decreasing behavior with $\alpha$. Notice how $T_{\text{BKT}}$ approaches $T_{\text{BCS}}$ with increasing $\alpha$.
For larger values of $U$, the superfluid weight increases in magnitude as contributions from the dispersive bands become relevant (see Fig.\,\ref{SWalp06}), and therefore $T_{\text{BKT}}$ increases overall.

\section{conclusions}\label{sec6}

In this work, we have studied the superconducting properties of the $\alpha$-$\mathcal{T}_3$ lattice model with on-site asymmetries, hosting an isolated quasi-flat band with tunable quantum metric. 
The sublattice-dependent superconducting order parameters were calculated self-consistently in mean-field theory as function of the interaction strength $U$, the hopping parameter ratio $\alpha$, and the electron filling per site $n_e$.
At quasi-flat band filling, the order parameters exhibit a rapid power law increase with 
$U$ due to the large density of states at such filling, in contrast to the usual exponentially small behavior \cite{Kopnin2011}.
Using the Kubo formula, we calculate the superfluid weight and decompose it into a conventional contribution determined by band derivatives, 
as well as in a geometric contribution involving state derivatives. 
Away from quasi-flat band filling the conventional contribution dominates as the bands are dispersive. However,
at partial quasi-flat band filling, the 
geometric contribution dominates over the conventional one for moderate values of $U$,
as in this regime the conventional contribution is determined by a vanishingly small band derivative.
At low enough $U$ where Cooper pairing occurs predominantly within the quasi-flat band, 
we find that the geometric contribution is determined mostly by the quantum metric of the quasi-flat band and increases linearly with $U$. 
We also find that the quantum metric of the quasi-flat band is enhanced as a function of $\alpha$, leading to an enhancement of the geometric contribution of the superfluid weight.
Correspondingly, the temperature-dependent superfluid weight and the BKT transition temperature are enhanced as $\alpha$ grows.

We trust our results would motivate the exploration of tunable quantum materials hosting flat bands as a source of rich quantum geometric phenomena and enhanced superconducting properties. 
 
\section{acknowledgments}

Supported by U.S. Department of Energy, Office of Basic Energy Sciences, Materials Science and Engineering Division.

\bibliography{biblio.bib}

\end{document}